\documentclass[prb,twocolumn,showpacs]{revtex4}
\usepackage{graphicx,psfrag,amssymb,amsmath}
\begin{document}
\title{The Wiedemann-Franz law in the SU(N) Wolff model}
\author{Bal\'azs D\'ora}
\email{dora@kapica.phy.bme.hu}
\affiliation{Department of Physics, Budapest University of Technology and 
Economics, H-1521 Budapest, Hungary}

\date{\today}

\begin{abstract}
We study the electrical and thermal transport through the SU(N) Wolff model with the use of bosonization. The Wilson 
ratio reaches unity as N grows to infinity. The electric conductance is dominated by the charge channel, and decreases 
monotonically with increasing interaction. The thermal conductivity enhances in the presence of local Hubbard $U$. The 
Wiedemann-Franz law is violated, the 
Lorentz number depends strongly on the interaction parameter, which can be regarded as a manifestation of spin-charge 
separation.

\end{abstract}


\maketitle

%\section{Introduction}

The Wiedemann-Franz law is one of the basic properties of a Fermi liquid\cite{abrikosov}. Simply stated, it 
reflect the 
fact, that the 
ability of quasiparticles to carry charge is the same as to transport heat. A possible breakdown of this relation is 
interpreted in terms of spin-charge separation. Spinons can carry heat in much the same way as charges do, and 
contribute to heat transport. On the other hand, they fail in electric transport, while charge excitations contribute to 
both. Another explanation requires inelastic scattering.

This phenomenon is demonstrated in the SU(N) Wolff model\cite{wolff,schlottmann,mattis,dorawolff}. It consists of N 
species of electrons 
interacting with each 
other only at a single site. The model is studied with Abelian bosonization\cite{nersesyan} for arbitrary N. 
We consider fermions with SU(N) spin index or alternatively
we take the ensemble of spin and orbital degrees of freedom into account by the N index\cite{szirmai,assaraf}.
 These additional
degrees of freedom can be realized through orbital degeneracy, for
example, as in Mn oxides\cite{imada}. As a result, the additional degrees of freedom can be called flavour
or color
index.
The Wolff model is one of the simplest impurity models, where electron correlation is still present\cite{wolff,mattis,zhang}.
Also it is the basis of studying the effect of Coulomb interaction on resonant tunneling through a single
quantum level\cite{ng,oguri}.

The Wilson ratio approaches unity as N grows, it
crosses 
over to non-interacting (mean-field) behaviour as N grows to infinity, a phenomenon characteristic to SU(N) 
models\cite{witten,coleman}.
The electric conduction involves only the charge sector, and increasing interaction increases the resistivity.
Heat is transported by both 
spin and charge excitation, and heat conduction is favoured by any finite value of $U$, regardless to its sign.  
The respective conductivities depend strongly on the local interaction, leading to the violation of the 
Wiedemann-Franz law even in the large N limit. This suggests that spin-charge separation is the origin of this 
breakdown. 

The Hamiltonian describing N different species of electrons interacting
only at the origin is given by:
\begin{gather}
 H=\sum_{m=1}^N\left[-iv\int\limits_{-L/2}^{L/2} dx
\Psi^+_m(x)\partial_x\Psi_m(x)+\right.\nonumber \\
E:\Psi^+_m(0)\Psi_m(0):+\nonumber \\
+\left.\frac U2
\sum_{n=1,n\neq m}^N:\Psi^+_m(0)\Psi_m(0)::\Psi^+_n(0)\Psi_n(0):\right],
 \label{hamilton}
\end{gather}
and only the radial motion of the particles is accounted for by chiral (right moving) fermion 
fields\cite{affleck2}.
The model can be bosonized via\cite{nersesyan,delft}
\begin{equation}
\Psi_m(x)=\frac{1}{\sqrt{2\pi\alpha}}e^{i\sqrt{4\pi}\Phi_m(x)},
\end{equation}
and after introducing charge and spin fields as\cite{assaraf}
\begin{gather}
\Phi_c(x)=\frac{1}{\sqrt{N}}\sum_{m=1}^N\Phi_m(x),\\
\Phi_{n,s}(x)=\frac{1}{\sqrt{n(n+1)}}\left(\sum_{m=1}^n\Phi_m(x)-n\Phi_{n+1}(x)\right),
\end{gather}
$n=1$\dots$N-1$, the Hamiltonian separates into different sectors: the spin sector is described by $N-1$ identical
decoupled bosonic modes, and the charge sector transforms into a similar
massless bosonic mode as
\begin{gather}
H_{s}=\sum_{m=1}^{N-1}\left[v\int\limits_{-L/2}^{L/2} dx 
(\partial_x\Phi_{m,s}(x))^2-\frac{U}{2\pi}(\partial_x\Phi_{m,s}(0))^2\right],\label{hamspin}\\
H_c=v\int\limits_{-L/2}^{L/2} dx (\partial_x\Phi_c(x))^2+E\sqrt\frac N\pi 
\partial_x\Phi_c(0)+\nonumber \\
+\frac{(N-1)U}{2\pi}(\partial_x\Phi_c(0))^2\label{hamcharge}.
\end{gather}
These can readily be diagonalized by introducing 
 pure bosonic representation of the $\Phi(x)$ fields as
\begin{equation}
\Phi(x)=\sum_{q>0}\frac{1}{\sqrt{2Lq}}\left(e^{iqx}b_q+e^{-iqx}b^+_q\right)e^{-\alpha
q/2},
\end{equation}
where $\alpha$ is the ultraviolet cutoff. Usually $x\in [-L/2,L/2]$, but
$L\rightarrow \infty$ in actual calculations.
The charge Hamiltonian is rewritten as
\begin{gather}
H_c=\sum_{q>0}\left\{vqb^+_qb_q+iE\sqrt{\frac{qN}{2L\pi}}(b_q-b^+_q)+\right.\nonumber\\
\left.+\frac{(N-1)U}{2\pi}\sum_{k>0}\frac{-\sqrt{qk}}{2L}(b_q-b^+_q)(b_k-b^+_k)\right\},
\label{hamboson}
\end{gather}
which describes spinless bosons scattered by $U$, and a source term
$E$. The latter can be transformed out by a linear shift of the bosonic
field, while the former can be considered exactly via Dyson equation.
Similar equations describe the spin sector.
Within the realm of bosonization, $U$ is restricted to $(-U_0/(N-1),U_0)$ with $U_0=\pi v/n_0$, $n_0$ is the average 
density per spin in the homogeneous case\cite{dorawolff}.

%\section{Wilson ratio}
As a first step to understand the response of our model, it is instructive to investigate the Wilson ratio. 
The Wilson ratio  characterizes to what extent the impurity interaction influences the conduction electron properties.
Using Ref. \onlinecite{dorawolff}, it is calculated for the SU(N) Wolff model as
\begin{equation}
R=\dfrac{\chi_{imp}}{\chi_0}\dfrac{\partial C_0/\partial T}{\partial C_{imp}/\partial T}=
1+\dfrac{U}{U_0+(N-2)U}.
\end{equation}
It interpolates smoothly between zero in the attractive case to $N/(N-1)$ in the repulsive case. Interestingly, the
latter value was calculated in the N-fold degenerate Coqblin-Schrieffer model\cite{hewson} as well.
This might suggest that the $U\rightarrow U_0$ limit of the Wolff model found by bosonization may describe the
$U\rightarrow\infty$ limit of the original model. The shortcoming of bosonization is to underestimate the strength of
interaction as it does for simple impurity scattering\cite{nersesyan}.

\begin{figure}[h!]
\psfrag{x}[t][b][1][0]{$U/U_0$}
\psfrag{y}[b][t][1][0]{$R$}
\centering{\includegraphics[width=7cm,height=7cm]{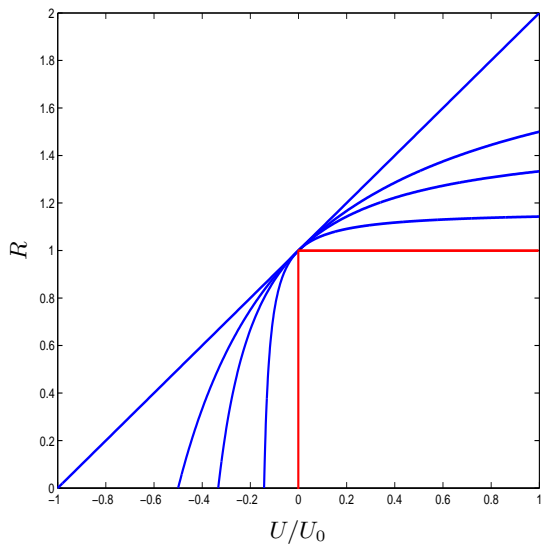}}
\caption{(Color online) The Wilson ratio of the SU(N) Wolff model is plotted for $N=2$, $3$, $4$, $8$ and $\infty$ 
(red) from top
to bottom. \label{wr}}
\end{figure}

As $N$ grows, $R$ approaches unity (i.e. the non-interacting limit). Indeed, the ground state of the
$N\rightarrow\infty$ limit of
impurity models is the mean-field solution\cite{hewson}, because fluctuations around the saddle-point are suppressed by
$1/N$, and vanish as $N$ reaches infinity. As a result, both quantities determining the Wilson ratio are proportional to 
the density of states at the Fermi energy\cite{dorawolff}, hence their ratio is 1.  The general behaviour of the Wilson 
ratio for various $N$'s is shown in Fig. \ref{wr}. The SU(2) case was also investigated in Ref. \onlinecite{zhang}.

The calculation of the conductivity can be carried out in several alternative ways. The most common one relies on the 
physical assumption that the whole voltage drop
occurs at the impurity site. The total number of electrons to the right and left of the impurity are $N_R$ and $N_L$, 
respectively. Then the current satisfying the continuity equation reads as\cite{nersesyan,izumida}
\begin{equation}
I=\frac e2\partial_t\left(N_R-N_L\right)=i\frac e2\left[H,N_R-N_L\right].
\end{equation}
From this, the linear conductivity can be calculated by the Kubo formula.

\begin{figure}[h!]
\psfrag{x}[t][b][1][0]{$\omega/W$}
\psfrag{y}[b][t][1][0]{$g(\omega)/g(0)$}
\includegraphics[width=7cm,height=7cm]{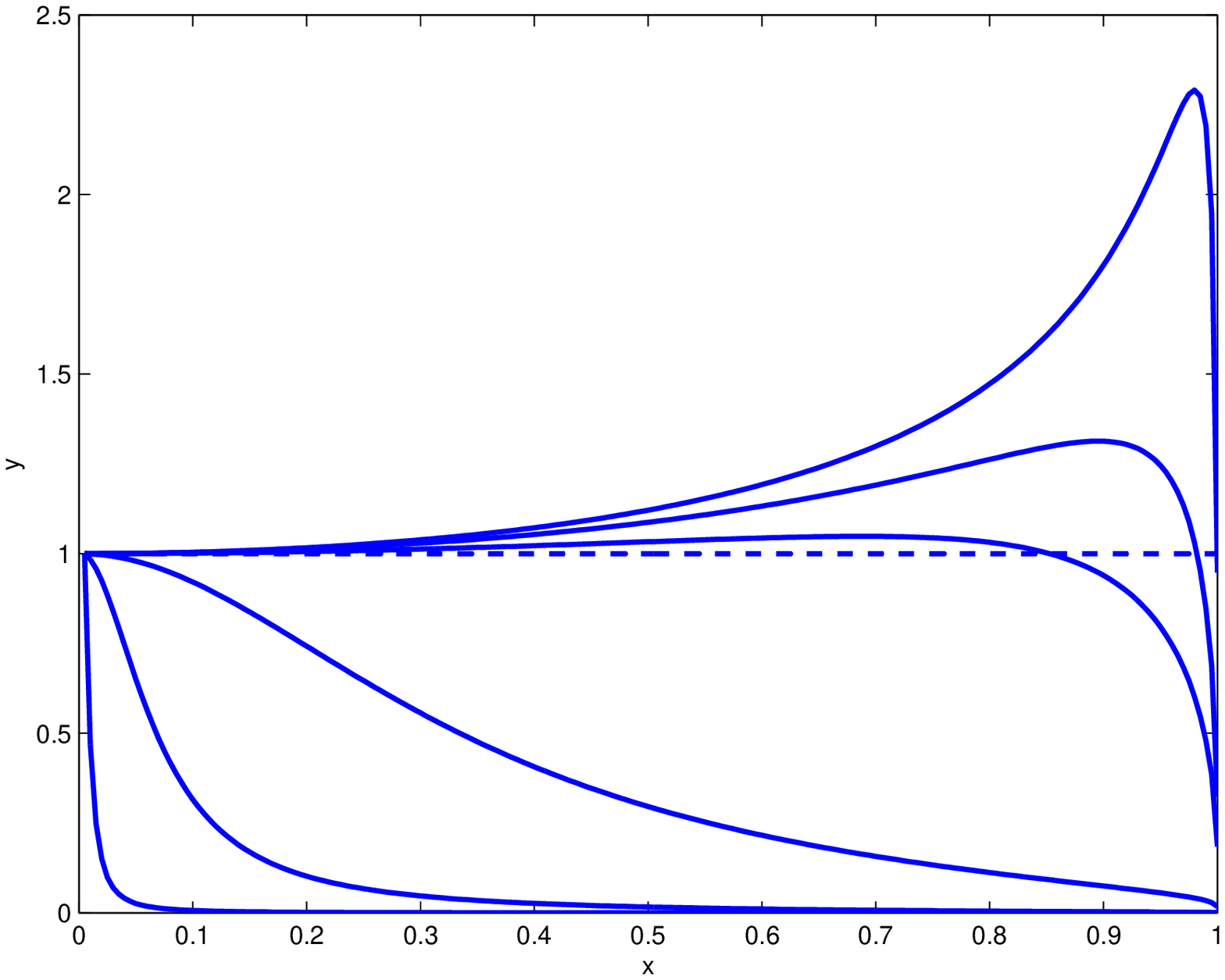}

\psfrag{x}[t][b][1][0]{$U/U_0$}
\psfrag{y}[b][t][1][0]{$g(0)2\pi/Ne^2$}
\vspace*{-6.7cm}\hspace*{-1.8cm}
\includegraphics[width=3.4cm,height=3cm]{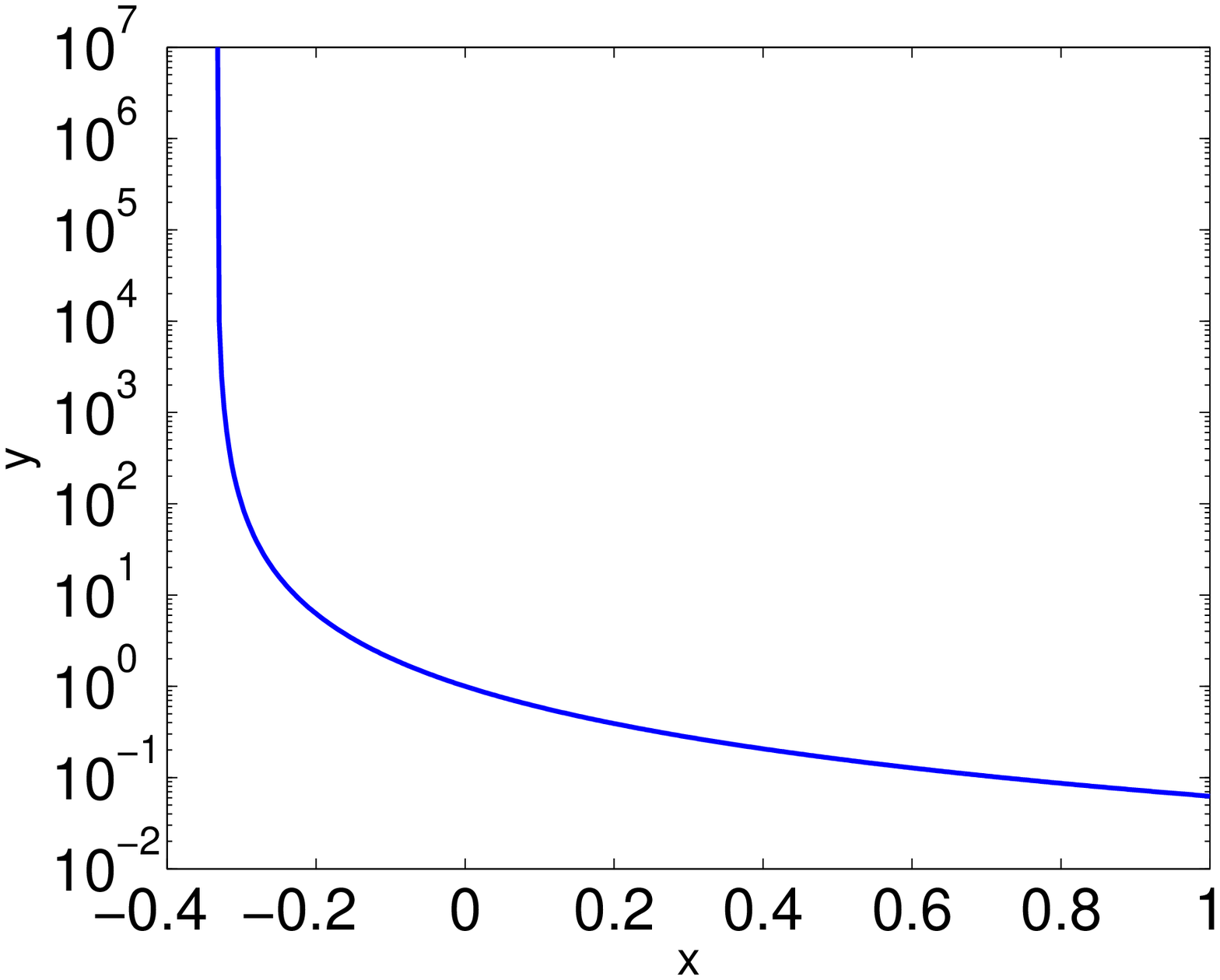}
\vspace*{3.7cm}
%\hspace*{2cm}
%\psfrag{x}[t][b][1][0]{$\omega/W$}
%\psfrag{y}[b][t][1][0]{$\phi(\omega)/\pi$}
%\includegraphics[width=7cm,height=7cm]{condphase.eps}
\caption{(Color online) The frequency dependent conductance is shown for $N=4$ for $U/U_0=-0.33$, $-0.3$, 
$-0.2$ below the dashed line indicating $U=0$ with increasing spectral weight. The other curves correspond to 
$U/U_0=0.2$, $0.5$ and $0.9$ with decreasing peak close to $\omega=W$, where $W$ is a sharp momentum space cut-off. The 
inset shows the dc conductance as a function 
of the Hubbard interaction.
\label{conduc}}
\end{figure}

On the other hand, one might think, that the conductivity at the impurity site is given by the autocorrelator of the 
current flowing through it. It turns out that these definitions of the current yield identical result, and we are in position
 to calculate the conductivity.
From the continuity equation ($\partial_t 
n+\partial_x j=0$), 
the local conductance at the 
impurity site is obtained, if we proceed following Ref. \onlinecite{fisher,oguri}. The local density is determined by 
\begin{equation}
n(x)=\sqrt{\dfrac{N}{\pi}}\partial_x\Phi(x),
\end{equation}
from which the local current operator is calculated as
\begin{equation}
j(x)=-e\sqrt{\dfrac{N}{\pi}}v\partial_x\Phi(x)=-evn(x).
\end{equation}
Only charge excitations participate in electric transport, spinons only transport heat, not charge, as will be 
demonstrated below. The expectation value of $\partial_x\Phi(x)$ is finite as seen in Ref. \onlinecite{dorawolff}, 
which is a natural consequence of the same chirality of all $N$ channels.
The current-current correlation function can be evaluated from the bosonic representation of the $\Phi(x)$ 
field 
through the equation of motion method. After solving the closed set of equations using Eq. \eqref{hamboson}, the 
frequency 
dependent ac conductance is evaluated 
as
\begin{equation}
g(\omega)=(ev)^2\textmd{Re}\dfrac{N\chi_0(\omega)}{i\omega(1+(N-1)U\chi_0(\omega))},
\end{equation}
where $\chi_0(\omega)$ is the local charge correlator per spin at $U=0$. Identical result was obtained for the 
charge correlator using the 
generating functional\cite{dorawolff}. 
It is shown in Fig. \ref{conduc} for $N=4$. Similar curves describe the $N\neq 4$ behaviour as well. 
Its dc part yields 
to 
\begin{equation}
g(0)=\dfrac{e^2}{2\pi}\dfrac{NU_0^2}{(U_0+(N-1)U)^2},
\label{dc}
\end{equation}
which decreases smoothly for repulsive interaction, but increases rapidly in the 
attractive sector. Throughout the calculations we set $\hbar=1$, which explains the $e^2/2\pi$ factor in front of the 
right hand side of Eq. \ref{dc}. By restoring the original units, this gets replaced by the universal channel conductance 
unit, $e^2/h$.
Qualitatively similar behaviour was identified in the 
conductance of a one dimensional
interacting electron gas (Luttinger liquid) through a weak link\cite{kane}. 
Repulsive interaction caused
perfect reflection, attractive one resulted in perfect transmission. 
Although the details are different,
our results exhibit the same phenomenon as seen from Eq. \ref{dc} and from the inset of
Fig. \ref{conduc}.
In the N$\rightarrow\infty$ limit, the dc conductance is zero for any finite $U$. The charge sector involved in electric 
transport is 
governed by Eq. \eqref{hamcharge}, where the scattering term ($\propto U$) grows with N, hindering any transport in the 
mean-field 
limit.

The ac 
conductance of an Anderson impurity was investigated in Ref. \onlinecite{sindel} in a different regime of the 
model, and hence with different conclusions from ours.

The heat transport can be studied similarly to the electric one. The heat-current operator is obtained from the energy 
conservation, following the steps outlined in Ref. \onlinecite{houghton,kanethermal}, and reads as
\begin{eqnarray}
J_Q&=&v^2\sum_{m=1}^N(\partial_x\Phi_m(x))^2\nonumber\\
   &=&v^2\left[(\partial_x\Phi_c(x))^2+\sum_{n=1}^{N-1}(\partial_x\Phi_{n,s}(x))^2\right].
\end{eqnarray}
The Hamiltonians in the different sectors commute with each other, hence their contributions can be determined 
independently. The thermal conductivity can be calculated from energy current-current correlation function given by 
\begin{equation}
\Pi(i\omega_n)=-\int\limits_0^\beta\textmd{d}\tau e^{i\omega_n\tau}\langle 
T_\tau 
J_Q(\tau)J_Q(0)\rangle,
\end{equation}
and the retarded response function is obtained after analytic continuation as
\begin{equation}
\frac{\kappa}{T}=-\frac{\textmd{Im}(\Pi(i\omega_n\rightarrow 
\omega+i\delta))}{T^2\omega}.
\end{equation}
Its calculation amounts to determine the $\langle T_\tau 
\partial_x(\Phi(0,\tau))^2(\partial_x\Phi(0,0))^2\rangle$ correlator. Possible pairings are of $\langle T_\tau
\partial_x\Phi(0,\tau)\partial_x\Phi(0,0)\rangle$ type in all 
sectors, which equals 
to the the charge 
and spin
correlation functions, obtained in 
Ref. 
\onlinecite{dorawolff}. The Hamiltonians describe scattering of free bosons on a localized "impurity", hence
 only self energy 
corrections can be taken into account (vertex corrections are absent). 
\begin{figure}[h!]
\psfrag{x}[t][b][1][0]{$U/U_0$}
\psfrag{y}[b][t][1][0]{$L/L_0$}
\centering{\includegraphics[width=7cm,height=7cm]{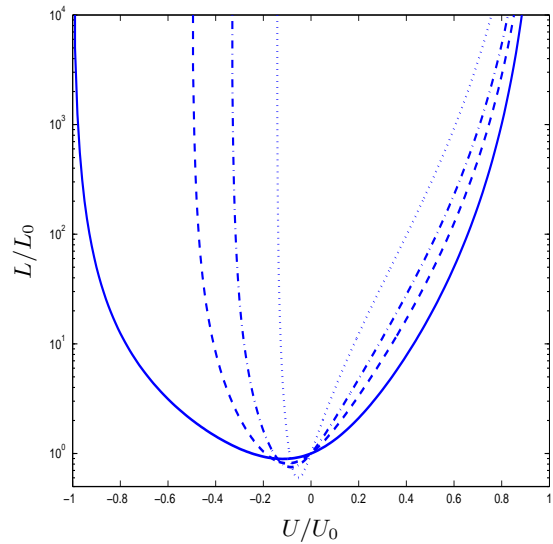}}
\caption{(Color online) The Lorentz number in the SU(N) Wolff model is plotted for $N=2$ (solid
line), $3$ (dashed
line), $4$ (dashed-dotted line), $8$ (dotted line) on a semilogarithmic scale. \label{lorentz}}
\end{figure}
The calculation can be carried out in a straightforward manner, and by taking all the channels into account,
we find
\begin{gather}
\frac{\kappa}{T}=\frac{1}{4T^3\pi}\int\limits_{-\infty}^\infty\textmd{d}x\frac{1}{\sinh^2\left(\beta
x/2\right)}\times\nonumber\\
\times\left[\left(\textmd{Im}G_c(x)\right)^2+(N-1)\left(\textmd{Im}G_s(x)\right)^2\right],
\end{gather}
where  $G_c(x)=v^2\pi\chi_0(x)/(1+(N-1)U\chi_0(x))$ and $G_s(x)=v^2\pi\chi_0(x)/(1-U\chi_0(x))$. As 
$T\rightarrow 0$, this can readily 
be 
evaluated to 
yield
\begin{equation}
\frac{\kappa}{T}=\frac{\pi}{6}\left(\frac{U_0^4}{(U_0+(N-1)U)^4}+\frac{(N-1)U_0^4}{(U_0-U)^4}\right).
\end{equation}
In the absence of $U$, it gives the pure result $\kappa=(\pi/6)NT$ (or $\pi^2k_B^2NT/3h$ upon restoring original units). 
In the presence of finite $U$, it increases 
regardless to the sign of the interaction. In other words, local Coulomb repulsion or attraction facilitates heat 
transport. From these, the Lorentz number reads as
\begin{gather}
L=\frac{\kappa}{T 
g(0)}=\frac{L_0}{N}\times\nonumber \\
\times\left[\frac{U_0^2}{U_0+(N-1)U)^2}+(N-1)\frac{U_0^2(U_0+(N-1)U)^2}{(U_0-U)^4}\right],
\end{gather}
where $L_0=\pi^2k_B^2/3e^2$ is the non-interacting result. For any finite $U$ and N, the Wiedemann-Franz law is 
broken, except a single value of attractive $U$ for any given N. This breakdown is a natural
consequence of spin-charge separation, as seen in Eqs. \eqref{hamspin} and \eqref{hamcharge}. 
Had we chosen $N=1$, the Wiedemann-Franz law would hold. As 
$U$ increases in the repulsive regime, it enhances strongly.
In the attractive case, a minimum value is reached for decreasing $U$ before the divergence at $-U_0/(N-1)$. 
For finite $T$, the Lorentz number decreases monotonically. Violation 
of the Wiedemann-Franz law was reported in one-dimensional electron gas (Luttinger liquid)\cite{kanethermal} with 
repulsive interaction. For any finite interaction, $L$ increased from the Fermi liquid value, similarly to our case. 
From these similarities in transport and in X-ray response\cite{dorawolff}, we conclude that the physical 
quantities 
evaluated within our model and its lattice version (Hubbard model) exhibit the same type of behaviour as a function of 
interaction.
Other failures of the Wiedemann-Franz law have been seen in mean-field theories as 
well\cite{houghton, sharapov}.
It is plotted in Fig. \ref{lorentz} for various N's.
Although the long time asymptotics of the Green's function is of Fermi liquid character, still spin-charge separation 
occurs on microscopic level, as seen in Eqs. \eqref{hamspin} and \eqref{hamcharge}, causing the breakdown of the 
Wiedemann-Franz law.
Similar phenomenon is expected to occur in the single- and multichannel realizations of the Kondo 
effect\cite{nersesyan}.

In conclusion, we have studied the electric and heat transport properties of the SU(N) Wolff model. As N grows to 
infinity, the systems crosses over to non-interacting behaviour. At the same time, the Wiedemann-Franz law remains 
broken due to spin-charge separation.

%\begin{acknowledgments}
This work was supported by the 
Magyary Zolt\'an postdoctoral
program of Magyary Zolt\'an Foundation for Higher Education (MZFK), and by the Hungarian
Scientific Research Fund under grant number OTKA TS049881.

%\end{acknowledgments}
\bibliographystyle{apsrev}
\bibliography{wboson}
\end{document}